\documentclass{article}

\usepackage{arxiv}

\usepackage[utf8]{inputenc} 
\usepackage[T1]{fontenc}    
\usepackage{hyperref}       
\usepackage{url}            
\usepackage{booktabs}       
\usepackage{amsfonts}       
\usepackage{nicefrac}       
\usepackage{microtype}      
\usepackage{lipsum}		
\usepackage{graphicx}
\usepackage{doi}
\usepackage{amsmath}
\usepackage{subfigure}

\def \ie {\emph{i.e.}}
\def \et {\emph{et al.}}

\def \ie {\emph{i.e.}}
\def \et {\emph{et al.}}
\def \ri { \mathbf{r}_i }
\def \vi { \mathbf{v}_i }
\def \Fi { \mathbf{F}_i }
\def \Fdij { \mathbf{F}^{D}_{ij} }
\def \Frij { \mathbf{F}^{R}_{ij} }
\def \Fcij { \mathbf{F}^{C}_{ij} }
\def \rij { \mathbf{r}_{ij} }
\def \vij { \mathbf{v}_{ij} }
\def \rmij { r_{ij} }
\def \rj { \mathbf{r}_j }
\def \vj { \mathbf{v}_j }
\def \eij { \mathbf{e}_{ij} }
\def \xij { \xi_{ij} }

\title{Dynamic spreading and infiltration of a molten sand droplet on a porous surface}

\date{} 					

\author{ \href{https://orcid.org/0000-0002-5212-1360}{\includegraphics[scale=0.06]{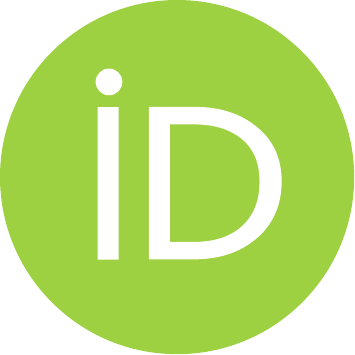}\hspace{1mm}Rahul Babu Koneru}\thanks{Corresponding Author} \\
	Department of Aerospace Engineering\\
	University of Maryland\\
	College Park, MD 20742 \\
	\texttt{rkoneru@umd.edu} \\
	\And
	{Garrett Foresman} \\
	Department of Chemistry and Biochemistry\\
    University of Maryland \\
    College Park, MD 20742\\
    \And
    {Alison Flatau} \\
    Department of Aerospace Engineering\\
	University of Maryland\\
	College Park, MD 20742 \\
    \And
    {Zhen Li} \\
    Department of Mechanical Engineering\\ Clemson University \\ Clemson, SC 29634
    \And
    {Luis Bravo, Muthuvel Murugan, Anindya Ghoshal} \\
    US Army Research Laboratory\\ Aberdeen Proving Ground, MD 21005
    \And
    {George Em Karniadakis} \\
    Division of Applied Mathematics\\ Brown University\\ Providence, RI 02912
}

\date{}

\hypersetup{
pdftitle={Dynamic spreading and infiltration of a molten sand droplet on a porous surface},
pdfsubject={Fluid dynamics},
pdfauthor={Rahul Babu Koneru},
}

\begin{document}
\maketitle

\begin{abstract}
	Compared to smooth surfaces, droplet spreading on porous surfaces is more complex and has relevance in many engineering applications. In this work, we investigate the infiltration dynamics of molten sand droplets on structured porous surfaces using the multiphase many-body dissipative particle dynamics (mDPD) method. We carry out three-dimensional simulations with different equilibrium contact angles and surface porosities. The temporal evolution of the radius of the wetted area follows a power law, as in the case of a smooth surface. The infiltration rate on the other hand is dictated by the competition between spreading and capillary inhibition of the pores. Additionally, the temporal evolution of the droplet height and the contact angle on the porous surface is also presented.
\end{abstract}


\section{Introduction}
Flow through porous media is widespread in nature and in many industrial applications \cite{lohse2022fundamental,gambaryan2014liquids}. On a nonporous surface, the early-time spreading is dictated by the inertial-capillary forces until viscous dissipation at the contact line enforces an equilibrium. On a porous surface, the spreading dynamics are more complex due to the competition between the lateral motion of the contact line and capillary inhibition of the pores. This depends on the physical properties of the system and also the topology of the porous substrate. The spreading of liquid droplets on porous substrates has been studied widely using theoretical, experimental and numerical approaches \cite{munuhe2022modeling, das2018droplet, frank2012droplet}. The Lattice Boltzmann (LB) simulations of Frank \et\ \cite{frank2012droplet} and fully-resolved volume-of-fluid simulations by Das \et\ \cite{das2018droplet} provide a great insight in to the dynamics of wetting and infiltration on porous surfaces for low viscosity liquids. Typically in most theoretical studies, the droplet is assumed to retain the spherical cap shape and the infiltration is treated using the Darcy's law. However, there is a substantial body of evidence which indicate a non-spherical shape of the droplet cap based on the topology of the surface textures and requires a detailed investigation at a pore level. 

Of particular interest is the spreading and infiltration of molten sand droplets, often referred to as CMAS, on the thermal barrier coating (TBC) of a gas turbine engine. The molten sand droplets infiltrate the porous TBC coating and upon solidification alter the thermal properties and cause mechanical failure. To develop CMAS-resistant TBCs, it is imperative to understand the underlying interfacial dynamics of the infiltration process.
In this work, three-dimensional (3D) numerical simulations are carried out using the many-body dissipative particle dynamics (mDPD) method to investigate the spreading and infiltration of a molten CMAS on porous substrates. The simulations are carried-out at two equilibrium contact angles: $\theta_{eq}=39^o$ and $70^o$. The porous surface is characterized by an isotropic arrangement of identical cylindrical pillars to represent the top-coat of a TBC. While a realistic CMAS deposition process is nonisothermal, the droplet is assumed to be isothermal at $1260\ \rm{^oC}$ in this work to isolate the effects of heat transfer.

\section{Numerical method} \label{sec:geq}
The interactions between mDPD particles are considered pairwise and their motion is governed by the Newton's second law of
motion~\cite{groot1997dissipative}. The position and velocity of the $i$-th mDPD particle is tracked using 
\begin{equation}
  \frac{d\ri}{dt} = \vi,\ m_i\frac{d\vi}{dt}= \Fi = \sum_{j \ne i} \Fdij + \Frij + \Fcij,
  \label{eq:geq}
\end{equation}
where $m_i$, $\ri$, $\vi$, and $\Fi$ denotes the mass, position, velocity, and force vector, respectively, of particle $i$. The total force $\Fi$ is the linear superposition of the dissipative ($\Fdij$), random ($\Frij$) and conservative forces ($\Fcij$) between neighboring particles $i$ and $j$, which are computed by
\begin{align}
  \Fcij &= A \omega_c(\rmij) \eij + B \left( \rho_i + \rho_j \right) \omega_d(\rij) \eij, \label{eq:fc}\\
  \Fdij &= -\gamma \omega_D(\rmij)\left( \eij \cdot \vij \right) \eij, \label{eq:fd} \\
  \Frij &= \beta \omega_R(\rmij) (dt)^{-1/2} \xij \eij. \label{eq:fr}
\end{align}
The pairwise interactions are governed by the weighting function $\omega_{(*)}$ which depends on the relative distance $\rmij=|\ri-\rj|$ between the particles $i$ and $j$. These interactions have a compact support and vanish beyond a cut-off distance $r_c$. The constants $\gamma$, $\beta$, $A$ and $B$ determine the strengths of each individual force. The relative velocity between a pair of particles is given by $\vij=\vi-\vj$. The forces between a pair of particles always lie along the line of centers $\eij=\rij/\rmij$. The Gaussian white noise $\xij$ is a random variable drawn from a Gaussian distribution with $\langle \xij(t) \rangle=0$ and $\langle \xij(t) \xi_{kl}(t^{\prime}) \rangle = \left( \delta_{ik} \delta_{jl}+\delta_{il} \delta_{jk} \right) \delta(t-t^{\prime})$ where $\delta_{ij}$ is the Kronecker delta and $\delta(t-t^{\prime})$ is the Dirac delta function. The system of stochastic ODEs is integrated in time using a modified velocity-Verlet (MVV) method.
The fluctuating-dissipation theorem (FDT)\cite{groot1997dissipative} leads to the following relation between the weights and the strengths of these two forces \begin{equation}
  \omega_D = \omega_R^{2},\ \beta^2 = 2\gamma k_B T,
  \label{eq:fdt}
\end{equation}
where $\omega_D = \omega_R^2 = \left( 1-\rmij/r_D \right)^s$ with $s=0.5$, $r_D=1.45$ and the non-dimensional temperature $k_B T=1.0$ where $k_B$ is the Boltzmann constant. The local density $\rho$ of a particle is computed as weighted sum of its neighbors \ie\ $\rho_i = \sum \omega_{\rho}\left( \rmij \right)$. For a more detailed discussion see \cite{koneru2022quantifying}.

\subsection{Parameter Mapping}
In an mDPD simulation the fluid properties such as $\rho$, $\sigma$ and $\nu$ are computed \textit{a posteriori} from mDPD simulations as opposed to the continuum-based methods where they are imposed directly. 
To accurately represent the properties of the fluid, the mDPD parameters have to be chosen carefully by mapping the mDPD parameters to the physical system of interest \ie\ molten CMAS. The physical properties of CMAS at $1260\ \rm{^oC}$ are \cite{naraparaju2019estimation}: density $\rho_l=2690$ $\rm{kg/m^{3}}$, surface tension $\sigma_l=0.46$ $\rm{N/m}$ and dynamic viscosity $\mu_l=3.6$ $\rm{Pa\cdot s}$. Following Arienti \et\ \cite{arienti2011many},
reference length [L], mass [M] and time [T] scales are computed to non-dimensionalize the system. The reference quantities are defined in terms of the reference units as
\begin{equation}
  \rho_{ref}=\frac{[M]}{[L]^3};\ \sigma_{ref}=\frac{[M]}{[T]^2};\ \nu_{ref}=\frac{[L]^2}{[T]}.
  \label{eq:ref}
\end{equation}
Following Koneru \et\ \cite{koneru2022quantifying}, the kinematic viscosity ($\nu$) and the surface tension ($\sigma$) of the mDPD system are computed using the doubly periodic Poiseuille flow and the thin-film method. The physical and the reference quantities obtained are tabulated in Table \ref{tab:units}.
\begin{table}
  \centering
  \caption{\label{tab:units} Physical and reference properties of molten CMAS.}
\begin{tabular}{lcc}
\textbf{Property} &\textbf{Physical units} &\textbf{mDPD units} \\
\hline
Density, $\rho$ & 2690 $\rm{kg/m^3}$ & 6.74 \\
Dynamic viscosity, $\mu$ & 3.6 $\rm{Pa \cdot s}$ & 196.08 \\
Kinematic viscosity, $\nu$ & $1.33\times10^{-3}$ $\rm{m^2/s}$& 29.093 \\
Surface tension, $\sigma$ & 0.46 $\rm{N/m}$ & 9.287 \\
Length $[L]$ & $17.017\times10^{-6}$   $\rm{m}$ & 1.0 \\
Mass $[M]$ & $1.964\times10^{-8}~\rm{kg}$ & 1.0 \\
Time $[T]$ & $6.297\times10^{-6}~\rm{s}$ & 1.0 \\
\end{tabular}
\end{table}

\subsection{Simulation setup}
\begin{figure}
  \centering
  \includegraphics[width=0.6\textwidth]{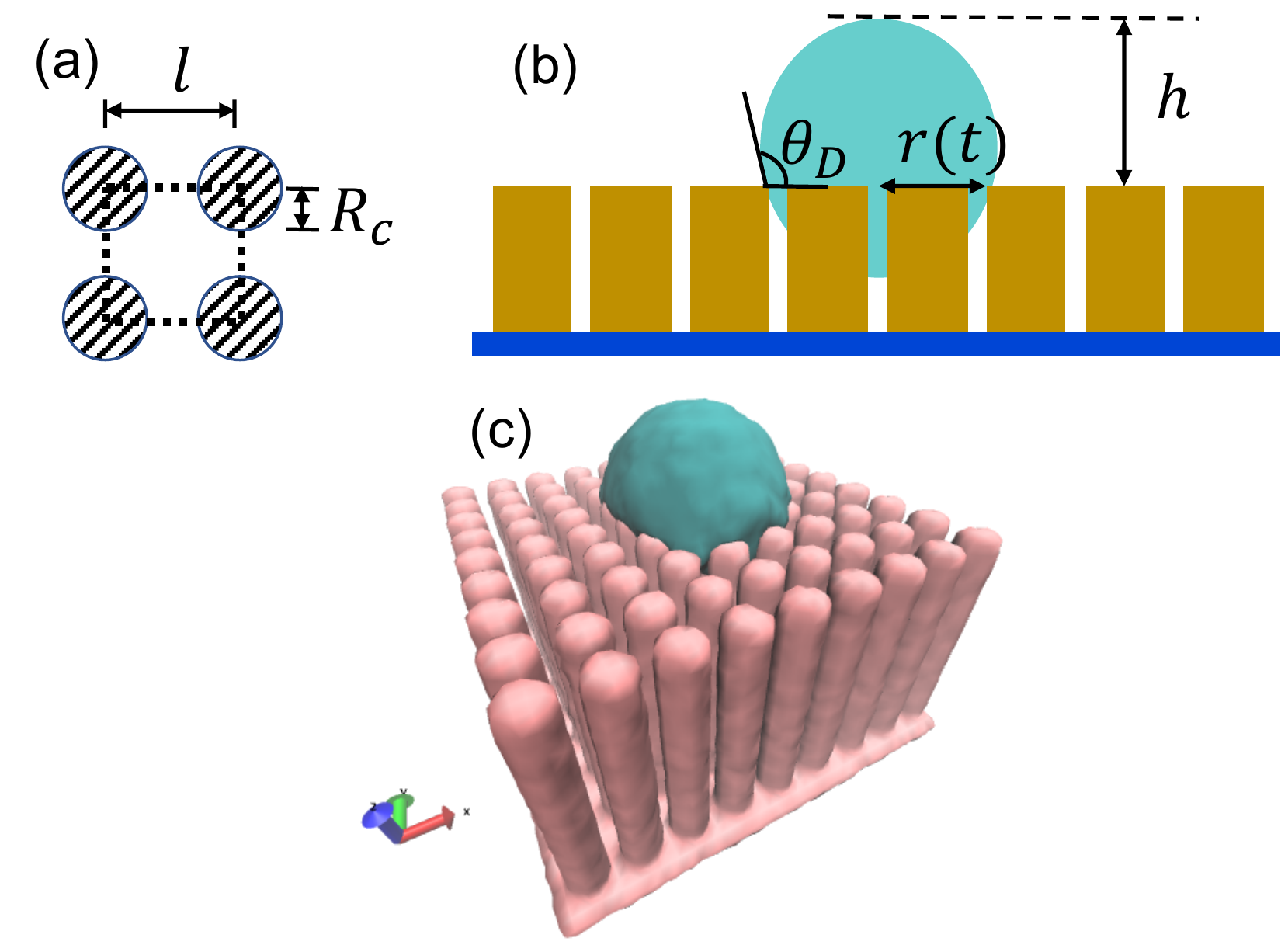}
  \caption{(a) A representative lattice of the surface porosity. The shaded circles denote the cross-section of the cylindrical pillars of radius $R_c$ separated by a distance $l$. (b) The contact angle, spreading radius and the droplet height measured with reference to the top of the porous surface. (c) Perspective view of the simulation depicting the droplet spreading on top of the pillars.}
  \label{fig:setup}
\end{figure}

The droplet and the porous wall are composed of randomly generated pre-equilibriated mDPD particles. The initial radius of the droplet is fixed at $R=10$ mDPD units in the simulations, which corresponds to 0.17 mm in physical units. 
The porous surface is comprised of an isotropic arrangement of cylindrical pillars. The void fraction of the surface $\phi_v = 1-\pi R_{c}^{2}/l$ is varied by changing the radius of the pillars while keeping the number of pillars identical. This way the droplet interaction with the cylindrical pillars remains consistent between different void fractions. A schematic of the cross-section of the porous surface is shown in Fig. \ref{fig:setup}.
In this work, simulations are carried out for $\phi_v = 0.35,\ 0.47\ \text{and}\ 0.58$. Various experimental studies ~\cite{kang2018high,naraparaju2019estimation} indicate a shallow equilibrium contact angle of molten CMAS. The mDPD parameters used in this work are identical to that in \cite{koneru2022quantifying} where the wetting behavior of molten CMAS was investigated on a nonporous surface.

\section{Results}
\begin{figure}
  \centering
  \includegraphics[width=0.85\textwidth]{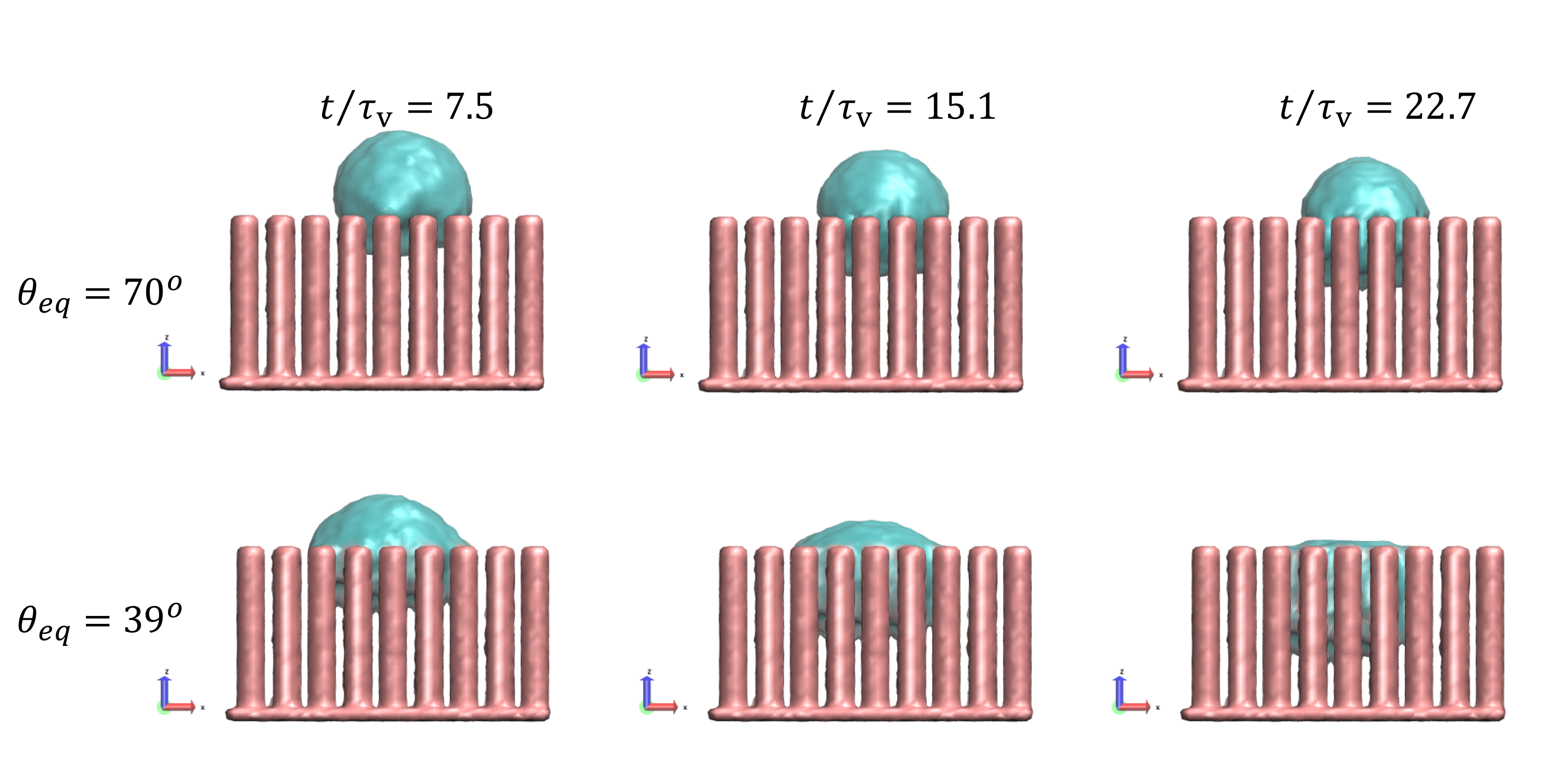}
  \caption{Droplet infiltration and spreading for different values of $\theta_{eq}$ at $\phi_v=0.58$.}
  \label{fig:evol_comb}
\end{figure}

The Ohnesorge number $Oh = \mu/\sqrt{\sigma \rho R}$ is typically used to identify the relative strength of the viscous and inertial force which for the molten CMAS droplet in this paper is 7.85. Given the large $Oh$ indicative of viscosity-dominated spreading, $\tau_v$ is used as the reference timescale. In Fig. \ref{fig:evol_comb}, the infiltration of the droplet on a surface with $\phi_v=0.58$ is shown at three different nondimensional times $t/\tau_v$ for droplets with an equilibrium contact angle of $\theta_{eq}=70^o$ and $39^o$ on a nonporous surface. In the case of $\theta_{eq}=39^o$, the droplet infiltrates completely within as short nondimensional time.  Due to combined lateral and normal inhibition of the droplet, the infiltration is highly nonhomogeneous and exhibits a departure from the spherical cap shape of the non-infiltrated portion of the droplet.

\begin{figure}
  \centering
  \subfigure
  {
    \includegraphics[width=0.6\textwidth]{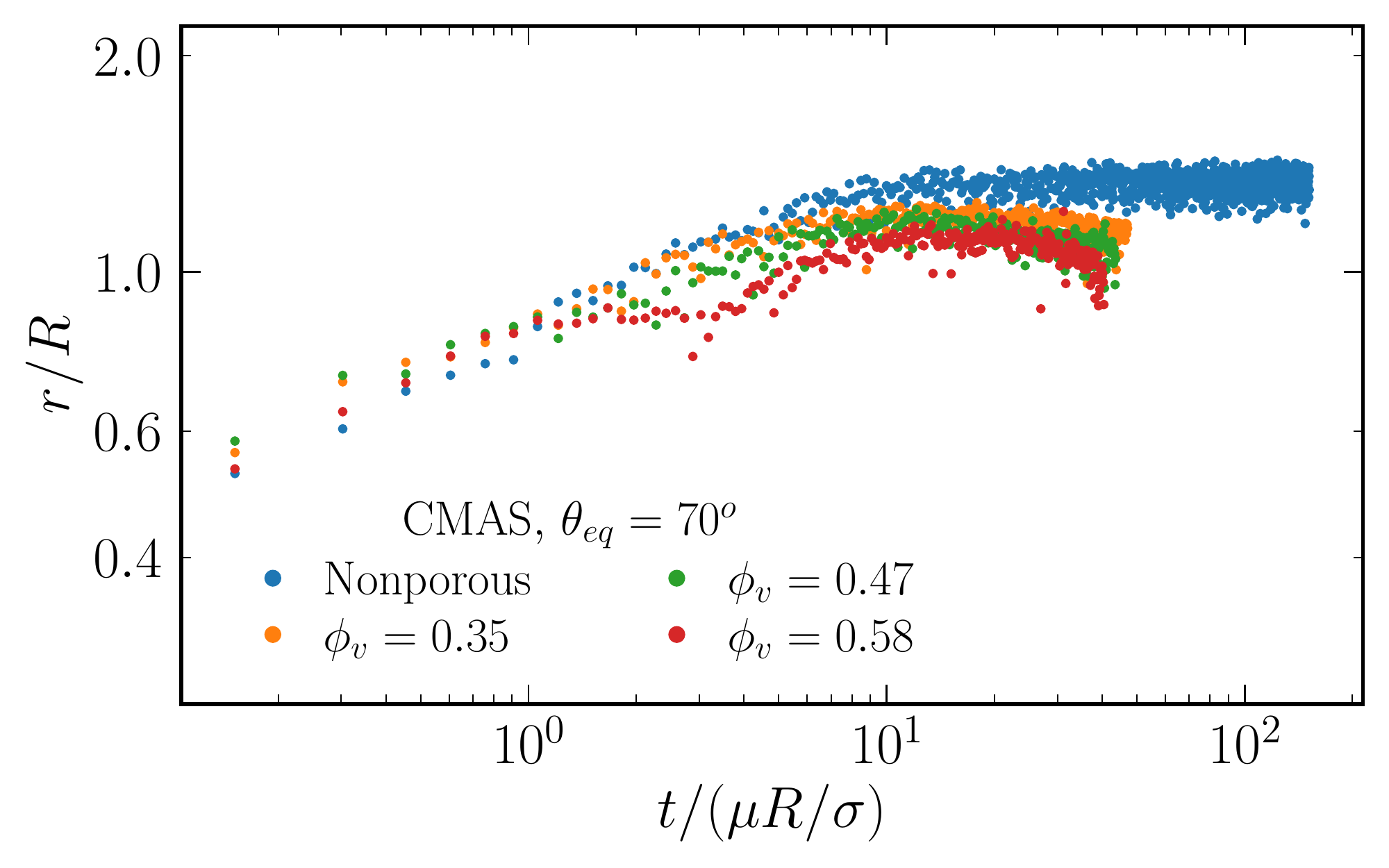}
  }
  \subfigure
  {
    \includegraphics[width=0.6\textwidth]{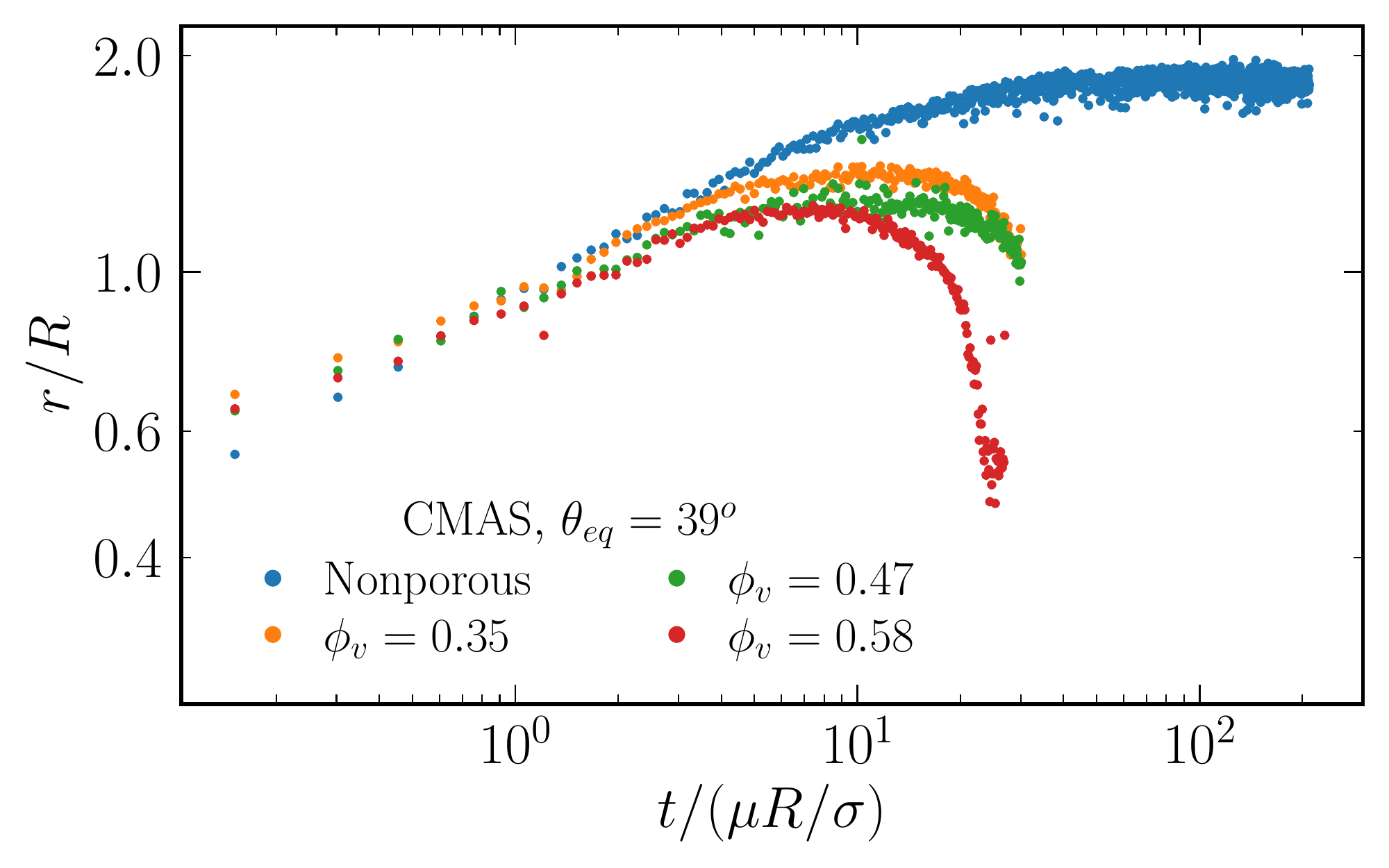}
  }
  \caption{(a) Nondimensional spreading radius of a molten CMAS droplet with (a) $\theta_{eq}=39^o$ and (b) $\theta_{eq}=70^o$.}
  \label{fig:rnd}
\end{figure}

In Fig. \ref{fig:rnd}, the nondimensional spreading radius, scaled by initial droplet radius $R$, is plotted as a function of nondimensional time which is scaled by $\tau_v$ for two configurations which produce an equilibrium contact angle $\theta_{eq}=70^o$ and $39^o$ on a nonporous surface. The droplet spreading is clearly characterized by an inertial regime which is soon followed by a reduction in the spreading radius due to infiltration. For a smaller $\theta_{eq}$, the rate of infiltration exceeds the rate of spreading much earlier compared to a larger $\theta_{eq}$. Compared to the case of $\theta_{eq}=70^o$, the transition at the end of the inertial regime is much sharper for $\theta_{eq}=39^o$ due to a larger contact area between the droplet and the porous surface. In \cite{koneru2022quantifying}, the inertial spreading rate of 0.26 was reported on a nonporous surface. Due to surface porosity, the spreading rate in the inertial regime is slightly reduced to a value between 0.18 and 0.21. 

\begin{figure}
  \centering
  \subfigure
  {
    \includegraphics[width=0.6\textwidth]{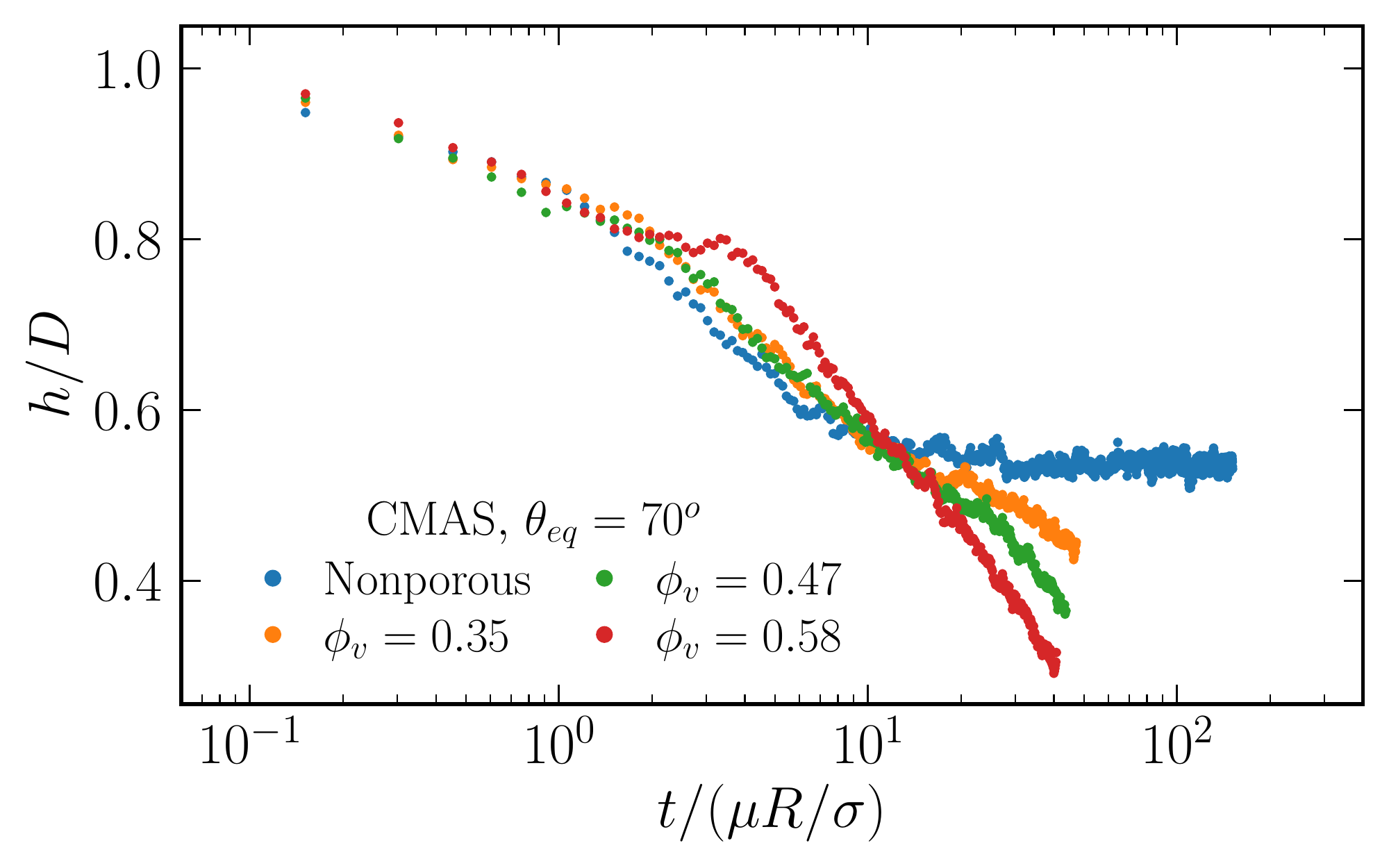}
  }
  \subfigure
  {
    \includegraphics[width=0.6\textwidth]{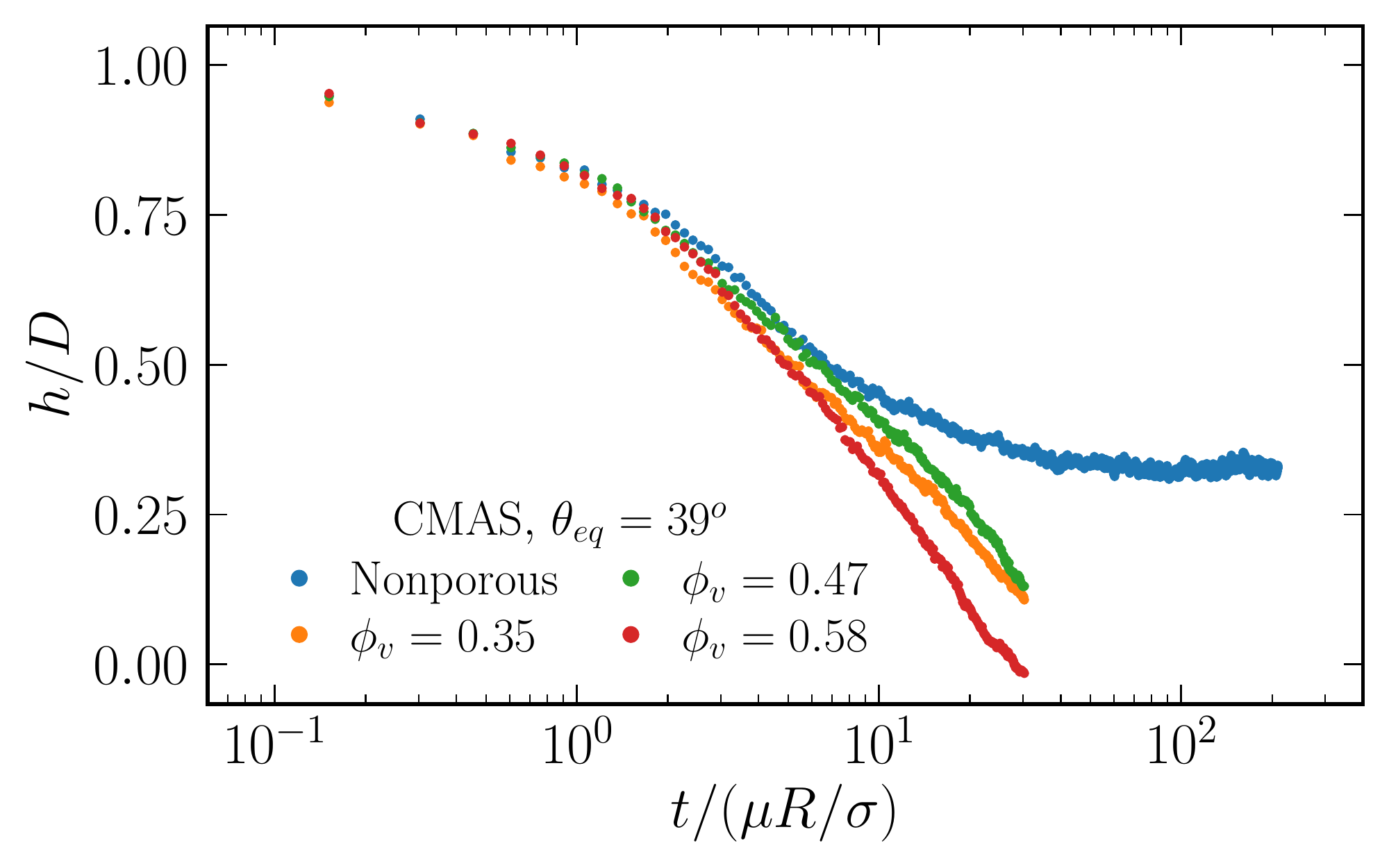}
  }
  \caption{(a) Nondimensional droplet height with (a) $\theta_{eq}=39^o$ and (b) $\theta_{eq}=70^o$.}
  \label{fig:htnd}
\end{figure}

\begin{figure}
  \centering
  \includegraphics[width=0.6\textwidth]{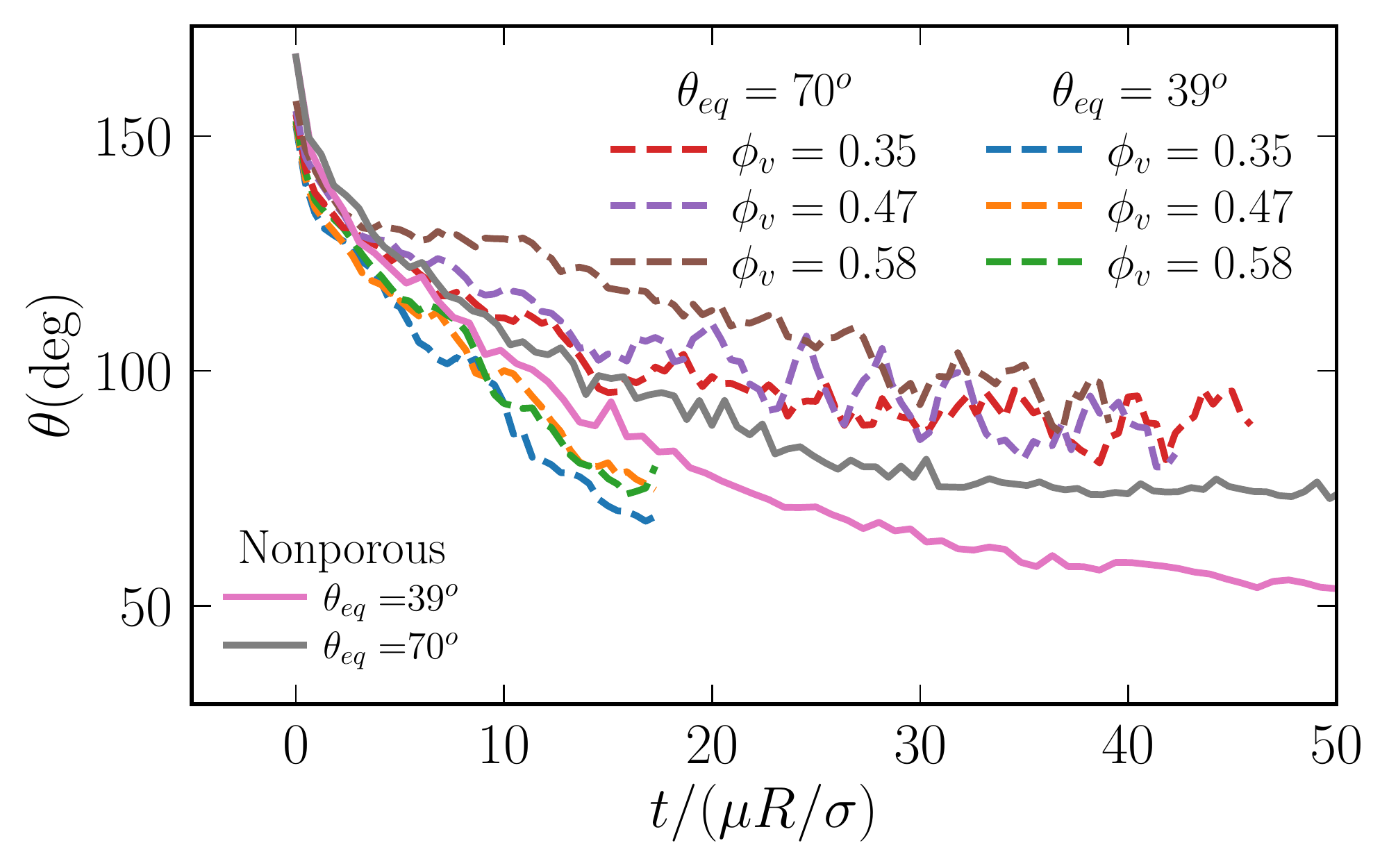}
  \caption{Temporal evolution of the apparent contact angle for $\theta_{eq}=39^o$ and $\theta_{eq}=70^o$.}
  \label{fig:theta_comb}
\end{figure}

The nondimensional droplet height measured from the top of the cylindrical pillars is plotted as a function of nondimensional time in Fig. \ref{fig:htnd}. For $\theta_{eq}=70^o$, the decrease in the droplet height is very similar for all the surface porosities until about $t/\tau_v=2$ after which the case with $\phi_v=0.58$ diverges. The drop height on the porous surface crosses over the curve on a nonporous surface at about $t/\tau_v=10$. This crossover was also reported by Frank \et\ \cite{frank2012droplet}. In the case of $\theta_{eq}=39^o$, this cross-over happens much earlier at about $t/\tau_v=2$ and a continuous and complete infiltration is observed. Since the spreading is driven by the inertia at early time, the differences in the drop height are minimal. In the later stages when capillary inhibition is dominant, the drop height decreases faster for the porous surface with the rate of decrease of droplet height increasing with increasing porosity in agreement with Washburn's law \cite{washburn1921dynamics}. 
The apparent contact angle of the droplet over the porous surface is also calculated by fitting an sphere through the mDPD particles present on the surface of the droplet. The temporal evolution of the apparent contact angle on the porous surface is shown in Fig. \ref{fig:theta_comb} along with the data from a nonporous surface. Interestingly, the porous surface behaves as a Cassie-Baxter surface for $\theta_{eq}=70^o$ wherein the apparent contact angle is larger on the porous surface than that on a nonporous surface. On the other hand, a Wenzel-like state is observed for $\theta_{eq}=39^o$ wherein the apparent contact angle is lower than that on a nonporous surface. Due to the large viscosity of molten CMAS, the droplet takes a very long time to reach an equilibrium on a nonporous surface. In this work, homogeneous wetting of the pore is observed wherein the porous surface is completely filled by the liquid.
\section{Conclusions}
In this work, we have investigated the spreading and infiltration of a molten CMAS droplet on a porous surface using two equilibrium contact angles and three different porosities. Due to the topology of the porous surface used in this work, the infiltrated liquid spreads around the cylindrical pillar resulting in a nonhomogeneous spreading. We also observe a departure from the spherical cap shape of the droplet. For a given porosity, a lower equilibrium contact angle induces a much faster drainage. Just like that on a nonporous surface, we observe an inertial spreading regime but with a reduced slope between 0.18 and 0.21. The competing effects between lateral spreading and capillary inhibition of the droplet produce interesting results. For both the contact angles, the droplet height reduces faster on the porous surface than that on a nonporous surface after a certain $t/\tau_v$. The crossover of the droplet height between the porous and nonporous occurs at much earlier for the smaller contact angle since a larger surface area of the droplet is in contact with the porous surface. We observe that the surface porosity has an opposite effect on the the evolution of the contact angle. Compared to a nonporous surface, the porous surface increases the apparent contact angle for a larger $\theta_{eq}$ whereas the apparent contact angle is reduced for a smaller $\theta_{eq}$.
%
\subsection*{Acknowledgements}
Authors A.F and R.B.K acknowledge the support received from the US Army Research Office Mathematical Sciences Division for this research through grant number W911NF1910225. L.B, A.G, M.M were supported by the US Army Research Laboratory 6.1 basic research program in propulsion sciences.  Z.L acknowledges the support from the NSF grant OAC-2103967 and the NASA project 80NSSC21M0153.  The authors gratefully acknowledge the resources and support provided by Department of Defense Supercomputing Resource Center (DSRC) through use of "Narwhal" as part of the 2022 Frontier Project, Large-Scale Integrated Simulations of Transient Aerothermodynamics in Gas Turbine Engines. The views and conclusions contained in this document are those of the authors and should not be interpreted as representing the official policies or positions, either expressed or implied, of the U.S. Army Research Laboratory or the U.S. Government. The U.S. Government is authorized to reproduce and distribute reprints for Government purposes notwithstanding any copyright notation herein.

\newpage
\bibliographystyle{unsrt}
\bibliography{References}





\end{document}